\documentstyle[multicol,prl,aps,psfig]{revtex}
\begin{document}
\tighten
\title{Tetragonal states from epitaxial strain on metal films}
\author{P.~M. Marcus$^1$ and P.~Alippi$^2$}
\address{$^1$ IBM Research Division, T.J. Watson Research Center,
Yorktown Heights, NY 10598, USA}
\address{$^2$ Fritz-Haber Institut der Max Planck Gesellschaft, \\ 
Faradayweg 4-6, D-14195 Berlin-Dahlem, Germany}
\maketitle
\begin{abstract}
The tetragonal states produced by isotropic pseudomorphic epitaxial
strain in the (001) plane on a tetragonal phase of a
crystal are calculated for V, Ti, Rb,
Li, K, Sr from first-principles electronic theory.  It is shown that
each metal has two tetragonal phases corresponding to minima of the
total energy with respect to tetragonal deformations,
hence are equilibrium phases, and that the equilibrium
phases are separated by a region of inherent instability.  The
equilibrium phase for any strained tetragonal state can thus be uniquely
identified.  Lattice constants and relative energies of the two phases
and the saddle point between them are tabulated, as well as the
tetragonal elastic constants of each phase. 
\end{abstract}

\noindent

\begin{multicols}{2}

\section{INTRODUCTION}

Metal crystals in the body-centered tetragonal
structure are known to have two total-energy minima as functions of
tetragonal lattice constants $a$ and $c$~\cite{1,2,3}.  The minima are 
appropriately called equilibrium metallic phases, since they persist
without applied stress and are stable under small tetragonal
deformations.  Previous work by the authors~\cite{4} defined and
discussed epitaxial Bain paths (EBP), which are sequences of
tetragonal states that include the equilibrium tetragonal
phases. These paths give the strained tetragonal states produced by
isotropic epitaxial strain, i.e., isotropic two-dimensional biaxial or
in-plane strain, imposed on the (001) planes of the equilibrium
phases.  In the previous work the EBP for V, Co, and Cu were found
using first-principles total-energy calculations and the EBP were
compared with the paths produced by {\it uniaxial} stress on the
phases, which were called in that work uniaxial Bain paths. 

The EBP were shown to be directly useful in interpreting the bulk
structure of epitaxial films determined, for example, by quantitative
low-energy electron diffraction (QLEED). Comparison of the measured
structure of a film with the states on the EBP identifies the phase
from which the film is produced by the epitaxial strain.  Thus a film
of Co on Cu(001) was shown to be strained fcc Co, but a film of Co on
Fe(001) was shown to be strained body-centered tetragonal (bct) Co, a
metastable phase of Co, whereas bcc Co was shown to be unstable.

In the tetragonal plane, whose coordinates are tetragonal lattice 
constants, the EBP is a continuous path that passes through the two
phase points at the energy minima and through the saddle point of
energy between the two minima. It was shown that between the minima a
segment of the EBP exists which includes the saddle point, but not the
minima, that consists of inherently unstable states.  More generally,
strained states of each phase, not just those on the EBP, were shown
to be separated by a region of inherently unstable tetragonal states,
so that an observed strained tetragonal state has a connection to just
one equilibrium phase through stable but constrained states.

The present work gives the EBP for six metals based on the published
first-principles calculations for tetragonal structures by Sliwko, Mohn,
Schwarz and Blaha~\cite{3}.  The calculation procedures for finding
the EBP, the contours of constant energy and the unstable region are
described in Sec.II.  The results are described in Sec.III with two
tables and four figures.  Section IV discusses why the EBP is useful,
the significance of the unstable region, and notes defects and
generalizations of these calculations.

\section{CALCULATION PROCEDURES}

The calculations of total energy $E$. as a function of the tetragonal
lattice parameters $a$, the side of the square cross section, and
$c$, the height of the unit cell, used the power-series expansions 
given in Ref.\cite{3}, whereas the calculations in Ref.\cite{4} used
the full-potential APW program Wien95 directly. The power-series
expansions were fitted in Ref.\cite{3} to extensive first-principles
calculations in the local-density approximation (LDA) with Wien95. 
The expansions give $E$ within specified ranges of $c/a$ and volume
per atom $V=ca^2 / 2$ that include the minima and saddle point and
have the form 
\begin{equation}
E= \sum_{i=0}^n \sum_{j=0}^m A_{ij} (c/a)^i V^j.
\label{Eq1}
\end{equation}
The coefficients $A_{ij}$ are tabulated in Ref.\cite{3} to eight
significant figures and are available in electronic form from the
authors. In Eq.(\ref{Eq1}) $n$ is $5$ or $6$, $m$ is $3$ and $E$ is
obtained in the specified ranges to an accuracy stated to be better
than $0.01$ mRy~\cite{5}. Some comments on this stated accuracy are
made in Sec.IV.  A useful feature of the formula (1) is that
analytical formulas for the first and second derivatives of $E$ may be
readily derived. 

The EBP for each metal is found by calculating $E$, 
$\left( \partial E / \partial c \right)_a$, and 
$\left( \partial E / \partial a \right)_c$ as 
a function of $c$ at constant $a$, and locating the $c$ value for  
which $E$ has a minimum or, equivalently, locating
the zero of $\left( \partial E / \partial c \right)_a$. 
The minimum corresponds to the epitaxial film condition of zero 
normal stress on the (001) surface. As $a$ ranges over the structures
between the phase points, the EBP is traced by these minima of $E$ at
each $a$. Since $E$ and $\left( \partial E / \partial c \right)_a$ 
are evaluated easily, a dense grid of $c$ values permits simple 
interpolation to four significant figures for $c$,$E$, and 
$\left( \partial E / \partial a \right)_c$ at the minimum of $E$ for
any $a$. 

The two phase points and the saddle point correspond to stationary
points for $E$, hence are located by interpolating on the EBP itself
to find points where $\left( \partial E / \partial a \right)_c$ 
vanishes along with $\left( \partial E / \partial c \right)_a$. Two of
the three stationary points are always cubic points, since at a cubic
point if $\left( \partial E / \partial c \right)_a =0$ then also
$\left( \partial E / \partial a \right)_c=0$. The contours of constant
$E$ are similarly found by interpolating the desired $E$ in a
tabulation of $E(c)$ at values of $a$ over a range of $a$ that covers
the desired contour. 

Tetragonal elastic constants at any $a$ and $c$ may be defined by 
\begin{center}
\begin{eqnarray}
\label{Eq2}
\bar{c}_{11} & = &{ {a^2} \over {V} } { {\partial^2 E} \over {\partial
a^2} } = 2 (c_{11} + c_{12}) , \nonumber \\ 
\bar{c}_{13} & = &{ {ac} \over {V} } { {\partial^2 E} \over {\partial
a \partial c} } = 2 c_{13} , \\ 
\bar{c}_{33} & = &{ {c^2} \over {V} } { {\partial^2 E} \over {\partial
c^2} } = c_{33} \nonumber 
\end{eqnarray}
\end{center}
The $\bar{c}_{ij}$ differ from the usual elastic stiffness
coefficients $c_{ij}$ because the $\bar{c}_{ij}$ correspond 
to tetragonal deformations which maintain the square symmetry
$a_1=a_2$~\cite{6}. To separate $c_{11}$ from $c_{12}$ 
requires breaking tetragonal symmetry, but is not possible if $E$ is 
known only from the power series (1). However when the phase has cubic
symmetry $c_{11} = c_{33}$ and $c_{12}$ can be evaluated. In 
fact :$c_{12}$ can be evaluated in two ways, i.e., from 
$\bar{c}_{11}$ and $\bar{c}_{33}$ on the one hand, and from
$\bar{c}_{13}$ on the other hand. 
The correspondence of the two values is then a test of the
accuracy of the power series representation of $E$ as will be noted in 
Sec.IV.

A strained tetragonal state will in general be maintained by applied
in-plane and out-of-plane stresses determined by the derivatives 
$\left( \partial E / \partial a \right)_c$ and $\left( \partial E /
\partial c \right)_a$. However stability depends also
on a condition on the second derivatives of $E$ which states that the 
second-order differential of $E$ is always positive, i.e., that 
\begin{equation}
\delta^2 E = V \left[  
{1 \over 2} \bar{c}_{11} \left(  { {\delta a} \over {a} } \right)^2  
+ \bar{c}_{13} { {\delta} \over {a} } { {\delta c} \over {c} } 
+ {1 \over 2} \bar{c}_{33} \left(  { {\delta c} \over {c} } \right)^2
                      \right] 
\end{equation}
is greater than zero for all deformations $\delta a$ and $\delta c$.
Otherwise the structure would have a tetragonal deformation that 
lowers the energy, so that the structure cannot be maintained by applied
stresses.  The condition $\delta^2 E > 0$ is then a condition 
on the $\bar{c}_{ij}$, namely, 
\begin{equation}
D =  \bar{c}_{11} \bar{c}_{33} - \bar{c}_{13}^2 > 0.
\end{equation}
The lines along which $D=0$ can be calculated readily by finding
the $c/a$ at which $D=0$ for the function $E(c/a)$ at constant $V$ and
using a range of $V$ to follow the line; the analytical power series
for the second derivatives of $E$. obtained from (1) are convenient 
for the calculation.

The slope of the EBP at the phase points can be expressed directly in
terms of the elastic constants of each tetragonal or cubic phase,
volume $V_0 = c_0  a_0^2 /2 $, 
\begin{equation}
\left( { {d(V/V_0)} \over {d(c/a)} } \right)_{\rm EBP} =
- { {a_0} \over {c_0} } 
  { {(2 \bar{c}_{33} - \bar{c}_{13})}  \over {(\bar{c}_{33} +
\bar{c}_{13} )} } = 
- { {2a_0} \over {c_0} } 
  { {( \bar{c}_{33} - \bar{c}_{13})}  \over {(\bar{c}_{33} + 
2 \bar{c}_{13} )} }.
\label{Eq5}
\end{equation}
If the phase point has cubic symmetry, Eq.(\ref{Eq5}) simplifies, since
then $c_{11} = c_{33}$, $c_{12} = c_{13}$ and
\begin{equation}
\left( { {d(V/V_0)} \over {d(c/a)} } \right)_{\rm EBP} =
- { {2a_0} \over {c_0} }
{ {(\bar{c}_{11} - \bar{c}_{12})}  \over {( \bar{c}_{11} + 2
\bar{c}_{12} )} } = 
- { {2a_0} \over {c_0} } { {1-2\nu} \over {1+\nu} }
\label{Eq6}
\end{equation}
where $\nu$ is the Poisson ratio of the cubic phase,
$\nu = c_{12} / (c_{11} + c_{12} )$.  In the figures
$V_0$ is chosen as the volume of the equilibrium phase of lower $E$,
hence Eq.(\ref{Eq5}) or Eq.(\ref{Eq6}) applied to the other equilibrium 
phase has a factor of the ratio of the $V$ of the other phase to $V_0$
on the right sides of Eq.(\ref{Eq5}) and Eq.(\ref{Eq6}) to get the
slope of the plotted EBP. 

Equations (\ref{Eq5}) and (\ref{Eq6}) relate the linear elastic
approximation to the EBP of a phase directly to the elastic constants
for a tetragonal noncubic phase or a cubic phase respectively.  If the
elastic constants of the phase are known from experiment or theory,
the equations give the linear approximation to the EBP.  Then the
measured bulk structure of a strained epitaxial film can be compared
with the linear EBP to identify the equilibrium phase of that film.
This identification is especially interesting for noncubic tetragonal
phases, which are predicted to exist for all transition
elements~\cite{1,3}, but are always metastable. 
Hence they cannot be made macroscopically, but may be stabilized by
epitaxy, as was done in the case of bct Co~\cite{4}. Comparison of
Eq.(\ref{Eq5}) with Eq.(\ref{Eq6}) shows that Eq.(\ref{Eq5}) defines an
effective Poisson ratio for tetragonal phases.

\section{RESULTS}

The results of calculation with the procedures and formulas of 
Sec.II are given in four figures and two tables.  Figures 1 to 3 plot
the EBP of V, Ti, and Sr along with contours of constant $E$ on the
$c/a-V/V_0$ tetragonal plane, where $V_0$ is the volume per atom of
the more stable phase point; the positions of the two phase points and
the saddle point are marked. The composite Fig.4 plots the EBP of Rb,
Li, and K without the contour lines.  The corresponding coordinates
and the energy at each point referred to a zero at the more stable
phase are given in Table I. Plotted in Figs.1, 2 and 3 are five
contours of constant energy, i.e., two contours at $\delta E$ above
the two minima, the contours through the saddle point, and the
contours $\delta E$ above and below the saddle-point energy. The
values chosen for $\delta E$ depend on the energy scale for each
metal. The unstable region where $D < 0$, which includes the  
saddle point, is the region between the two lines of long dashes, and
is shown in all four figures.  Table I also gives the stationary
points of $E$ for tetragonal Rb computed by Milstein, Marschall and
Fang~\cite{7} from an empirical potential fitted to experiment. 
%
%
\begin{figure}[h]
\centerline{\hbox{\hspace{-2cm}
\psfig{figure=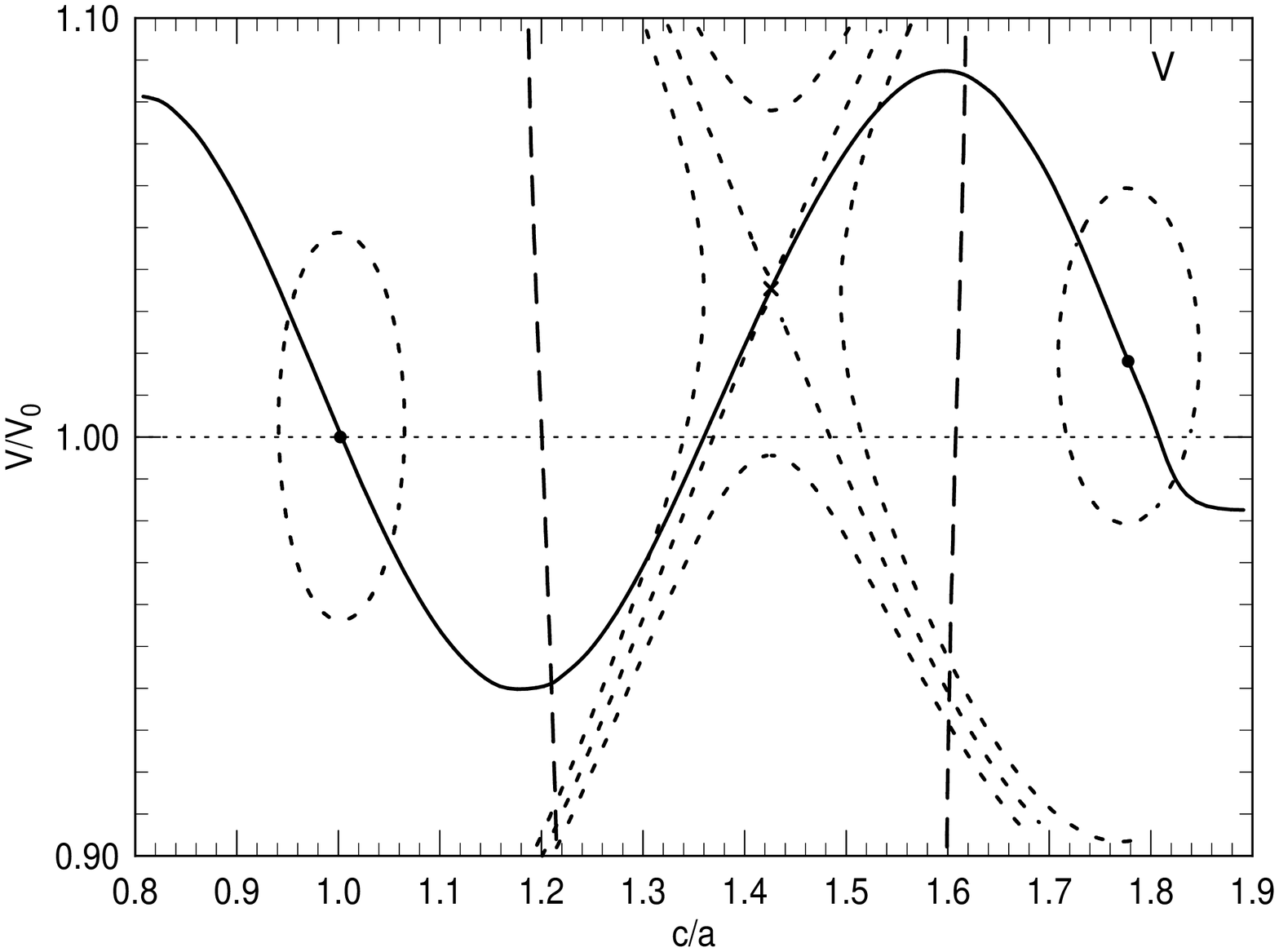,width=8cm,height=8cm}}}
\vspace{-1.5cm}
FIG. 1. EBP for V on the $c/a-V/V_0$ plane (full line), phase
point 1 bcc (full circle), volume $V_0 = 84.8 {\rm bohr}^3$, $E=0$; 
saddle point fcc (marked $\times$) $E=20.8$ mRy; phase point 2 
bct (full circle), $E=10.0$ mRy. Energy contours are drawn
(short dashes) at $\delta E = 1$ mRy above the minima, through
the saddle point and $\pm \delta E$ from the saddle point.  The unstable
region is between the lines of long dashes and includes the saddle
point. The coordinates for the phase and saddle points are in 
Table I.
\label{fig1}
\end{figure}

The tetragonal elastic stiffness constants $\bar{c}_{11}$ , 
$\bar{c}_{13}$, $\bar{c}_{33}$ are given in Table II at the phase points 
and also the elastic stiffness constants found in Ref.\cite{3} for the 
cubic phases of each metal. When the phase is cubic, the usual tetragonal 
elastic constants $c_{11}$ and $c_{12}$ (found in two ways) are
given; note that then $c_{11} = \bar{c}_{33}$.
By symmetry, $E$ is stationary at the cubic points $(c/a= 1, \sqrt 2)$
on the EBP, i.e., $\partial E / \partial a = \partial E / \partial c
=0$ at cubic points.  However the cubic points can be saddle points as
well as minima. The possible configurations have been classified in
Ref.\cite{7} in three cases, i.e., Case 1: minimum at bcc ($c/a=1$), saddle
point at $1 < c/a < \sqrt 2$, and minimum at fcc ($c/a=\sqrt 2$);
Case 2: minimum at bct ($c/a < 1$), saddle point at bcc ($c/a=1$),
minimum at fcc ($c/a= \sqrt 2$); Case 3: minimum at bcc ($c/a=1$),
saddle point at fcc ($c/a= \sqrt 2$), minimum at bct ($c/a > \sqrt
2$). Then Rb, K, Li, Sr are Case 1, Ti is Case 2 and V is Case 3.
%
%
\begin{figure}[h]
\centerline{\hbox{\hspace{-2cm}
\psfig{figure=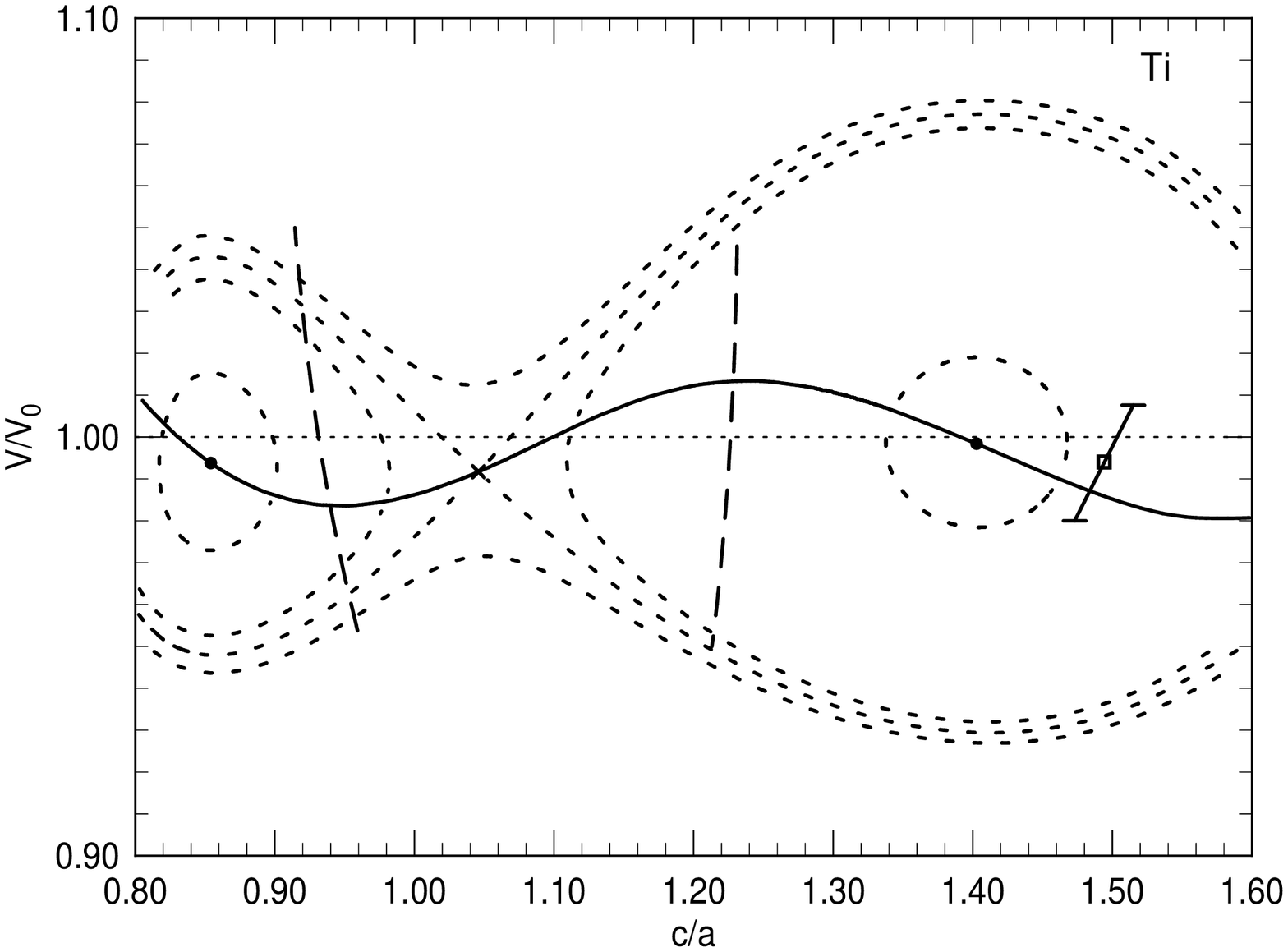,width=8cm,height=8cm}}}
\vspace{-1.5cm} 
FIG. 2. EBP for Ti, stationary points, contour lines and 
unstable region marked as in Fig. 1.  Phase point 1 is bct at $E=1.59$ 
    mRy; saddle point is bcc at $E=2.59$ mRy; phase point 2 
   is fcc at $E=0$ mRy, $V_0 = 108.1$ bohr$^3$; 
   $\delta E = 0.2$ mRy.  The measured strained bulk
  structure of epitaxial film on Al(001) is marked by the open square
  with error line \cite{8}.
\label{fig2}
\end{figure}
%
%
%
%
\begin{figure}[h]
\centerline{\hbox{\hspace{-2cm}
\psfig{figure=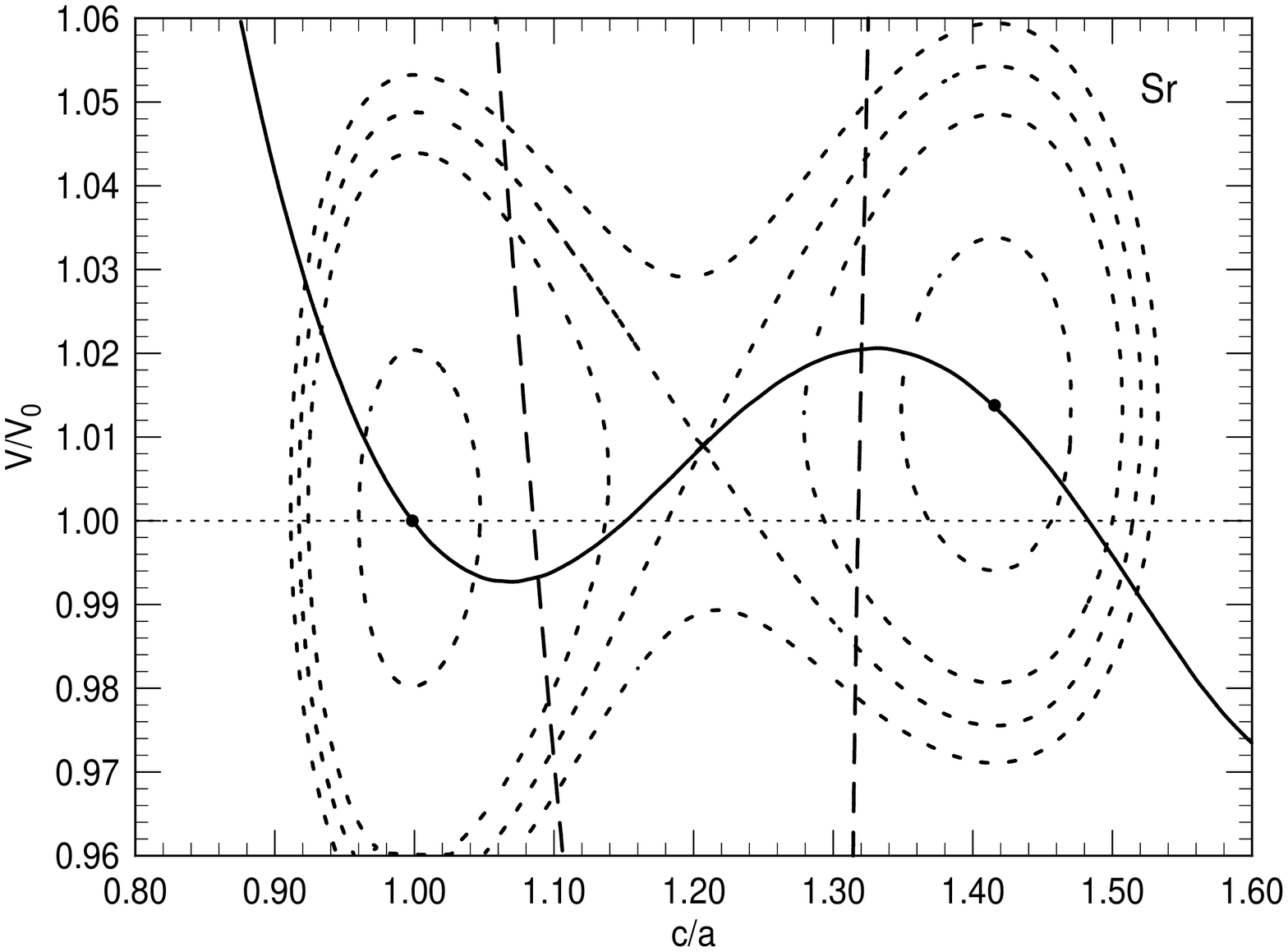,width=8cm,height=8cm}}}
\vspace{-1.5cm}
FIG. 3. EBP for Sr, stationary points, contour lines and unstable region
marked as in Fig, 1,  Phase point 1 is bcc at $E=0$ mRy; saddle
point is bct at $E=0.383$ mRy; phase point 2 is fcc at
$E=0.190$ mRy, $V_0 = 319.6$ bohr$^3$, $\delta E = 0.07$ mRy.
\label{fig3}
\end{figure}

The contour lines of constant $E$, which are vertically-oriented 
ellipses near each phase point, also appear in Ref.\cite{3} plotted on
the $c/a-V$ plane for each metal. The EBP and contour lines for $V$
are also in Ref.\cite{4}, where they are plotted on the $a/a_0 - V/
V_0$ plane. On this plane the contours are tilted and the bcc and fcc
positions on the EBP are reversed.  The contours on the $c/a-V$ plane
in terms of the deviations of $c/a$ and $V$ from the values at the
phase points are given by 
\begin{eqnarray}
E- E_{\rm min} & = &
c_1 \! \left( { {\delta c/a} \over {c/a} } \right)^2 \! \! + 
c_2 \! \left( { {\delta c/a} \over {c/a} } \right)
       \left( { {\delta V}   \over {V}   } \right) \! \! + 
c_3 \! \left( { {\delta V}   \over {V}   } \right)^2 , \nonumber \\
c_1 &=& (V/18) ( \bar{c}_{11} - 4 \bar{c}_{13} + 4 \bar{c}_{33} ),
\nonumber \\
c_2 &=& -(V/9) ( \bar{c}_{11} - \bar{c}_{13} -2 \bar{c}_{33} ), \\
c_3 &=& (V/18) ( \bar{c}_{11} + 2 \bar{c}_{13} +  \bar{c}_{33} ) \\
\nonumber  
\label{Eq7}
\end{eqnarray}

%
%
\begin{figure}
\centerline{\hbox{\hspace{-2cm}
\vspace{1cm}\psfig{figure=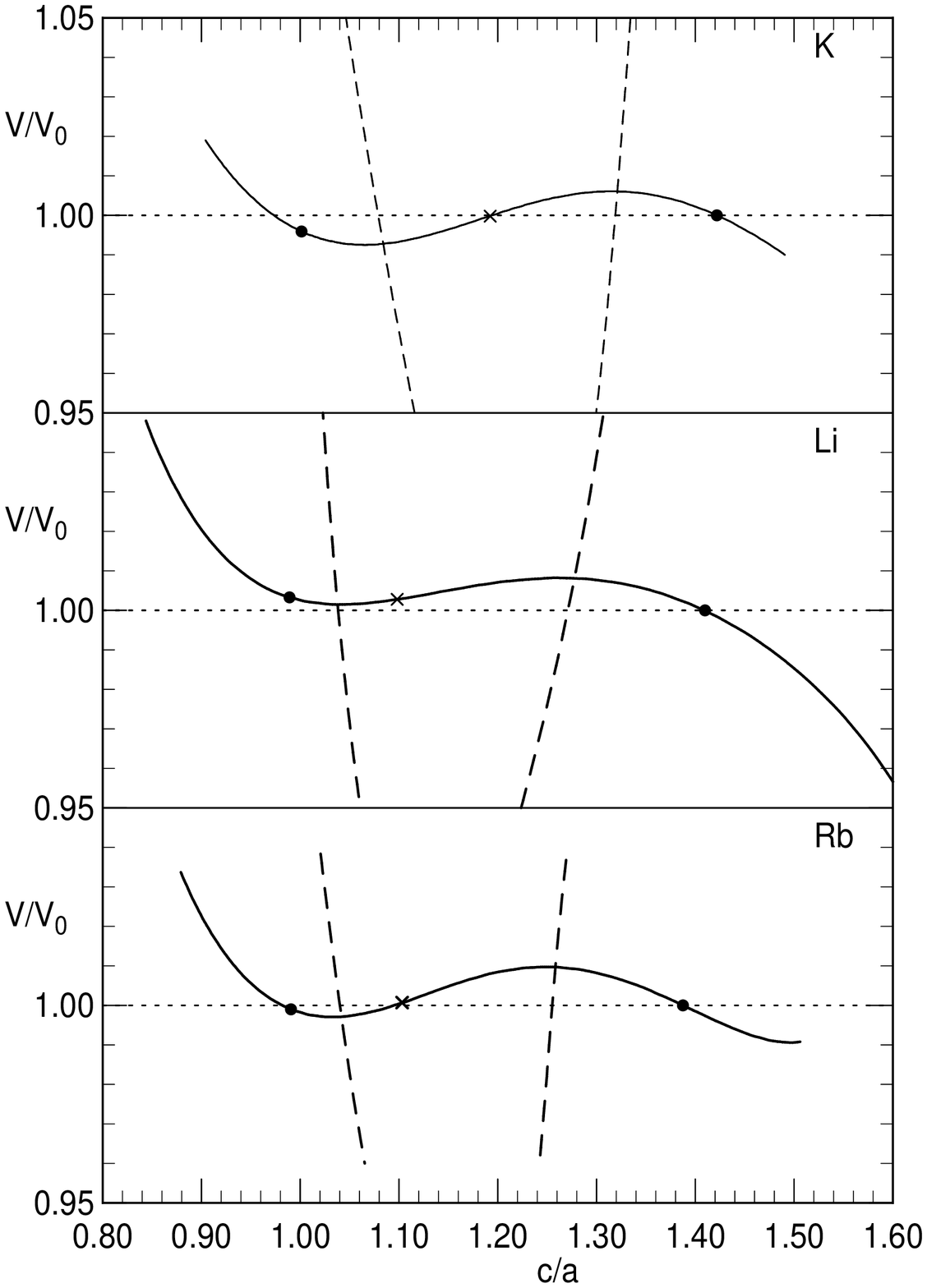,width=8cm,height=8cm}}}
  FIG. 4. EBP for Rb, Li, K, stationary points, and unstable region
  marked as in Fig. 1.  Phase point 1 for Rb is bcc at $E=0.135$
  mRy; saddle point is bct at $E=0.157$ mRy; phase point 2 is
  fcc at $E=0$ mRy, $V_0 = 518.0$ bohr$^3$. 
  Phase point 1 for Li is bcc at $E=0.144$ mRy; saddle point is
  bct at $E=0.161$ mRy; phase point 2 is fcc at $E=0$ mRy, 
  $V_0 = 127.7$ bohr$^3$.
  Phase point 1 for K is bcc at $E=0.013$ mRy; saddle point is
  bct at $E=0.089$ mRy, $V_0 =430.4$ bohr$^3$.
\label{fig4}
\end{figure}
For cubic symmetry $c_2$ vanishes; even for the bct phases of Ti and
V, $c_2$ is much smaller than $c_1$ and $c_3$, so the contours are
still nearly vertical. Formula (7) for the case of cubic symmetry 
with coefficients in terms of the $c_{ij}$ is given in Ref.\cite{2}. 

{\section{DISCUSSION}}

The principal result found here is the EBP between the phase points,
which shows the effects of isotropic epitaxial strain on equilibrium
phases.  The EBP provides a basic elastic response of a material in
tetragonal structure to a particular strain, one which is available
experimentally, including the interesting case of negative strain in the
plane.  The EBP are found here from first principles in a well-defined
approximation with errors of known magnitude, and include nonlinear
effects.
These EBP can then be compared directly with the strained
bulk structures determined by QLEED for epitaxial films.  This
comparison is illustrated for strained epitaxial Ti films on
Al(001)~\cite{8}, where the QLEED point and its error limits 
(from the uncertainty in the bulk value of $c$) 
are shown in Fig. 2 to agree well with the calculated EBP.  The 
$V_0$ used in evaluating $V/V_0$ for the QLEED point is the measured
hcp Ti value, $V_0^{\rm hcp} =119.2 {\rm bohr}^3$ which is close to
the fcc value. 

The presence on the EBP of an inherently unstable section separating
strained fcc Ti from strained bct Ti is an important result of the
theory.  Since the QLEED point within its error limit lies on the fcc
part of the EBP, the epitaxial film must be strained fcc Ti.  This
result is particularly interesting because fcc Ti does not appear on
the usual pressure - temperature phase diagram.  Note that the theory
gives directly the possible states of epitaxially-strained fcc Ti,
including any nonlinear elastic behavior of the crystalline phases.
This comparison of measured structure with the theoretical EBP
replaces the previous analysis, which assumed constant elastic
stiffness coefficients and attempted to estimate the elastic
coefficients of the cubic phases~\cite{8}.  The use of linear elastic 
relations for identification of the equilibrium phase is compared to
the use of the EBP for that identification in the case of Co in
Ref.\cite{9}, where tetragonal states and the EBP are plotted on the
$a-d$ plane, where $d$ is the layer spacing.

These first-principles calculations, which do not use empirical
information, have an advantage over empirical potentials fitted to
measurements, since these calculations are
as good for highly strained or even unstable states,
which are not accessible to measurement, as they are for slightly
strained equilibrium states.  Thus they
provide a test of calculations based on empirical
potentials, such as the potential used for Rb in Ref.\cite{7}.
Comparison in Table I shows that the $c/a$ values of the saddle point
differ by $10 \%$, that the minimum energy in Ref.\cite{7} is at the
bcc structure, rather than the fcc structure found here, and that the
energy separation of the equilibrium phases in Ref.\cite{7} has both a sign
and a magnitude different (smaller) from what is calculated here. 

The power series representation of the tetragonal energies shows some
defects, e.g., the cubic points deviate from $c/a=1$ or $\sqrt 2$. in
Table I, the two values of $c_{12}$ do not agree well in some cases. 
These defects are in the representation of the results of LDA
calculations.  The most serious defect is in the saddle point of Ti,
which is bcc, but the power series finds the saddle point at 
$c/a=1.05$ rather than $1$.  A recalculation of the energy $E$ of
Ti directly with \verb|WIEN95| finds that the power series has missed
an  asymmetry in $E$ around the saddle point which shifts the position
of the maximum.  The direct calculation finds the maximum at $c/a=1$ 
as it should be; it also verifies the minimum of fcc Ti at $c/a=1.40$.
The deviation from symmetry at fcc Ti and all other cubic phases is
thus no more than $1\%$, except for fcc Rb, where the deviation is
$1.7 \%$. 
In comparison with experiment all the calculated elastic stiffness
coefficients are too large by at least $10 \%$ and the volumes per
atom are too small by $5$ to $10 \%$.  These discrepancies from
experiment are defects of the assumptions of the band calculations,
i.e., of the LDA with semi-relativistic corrections. 

\end{multicols}
\begin{figure}
\refstepcounter{table}
\small{TABLE I. Tetragonal states stationary in energy. The parameter are 
$(c/a)_i$,$(V/V_0)_i$, $E_i$, $i=1,2,3$, in the tetragonal plane for 
states of stationary $E$; the stable phase point is $i=1$ or $3$
and the metastable phase point is then $i=3$ or $1$, respectively;
the saddle point is $i=2$.  Note that the tetragonal lattice
parameters can be found from
$ a = [2 V_0 (V / V_0 )/ (c/a)]^{1/3}$ and
$ c = a (c/a)$.  $V_0$ is the volume per atom of the more stable phase 
point in bohr$^3$ and $E_i$ is the energy in mRy with respect
to the energy of the more stable phase point.}

\begin{center}
\begin{tabular}{ccccccccccc}
\hline \hline
Metal & $V_0$   & $(c/a)_1$  & $(V/V_0)_1$  & $E_1$ & 
$(c/a)_2$  & $(V/V_0)_2$  & $E_2$ &
$(c/a)_3$  & $(V/V_0)_3$  & $E_3$ \\
\hline 
V  & $84.9$ & $1.00$ & $1.00$ & $0.00$ & $1.43$ & $1.04$ & 
$20.8$  & $1.78$ & $1.02$ & $10.0$ \\
Ti & $107.9$ & $0.85$ & $0.99$ & $1.59$ & $1.05$ & $0.99$ & 
$2.59$ & $1.40$  & $1.00$ & $0.0$ \\
Rb$^a$  & $518.0$ & $0.99$ & $0.999$ & $0.135$ & $1.10$ & 
$1.001$ & $0.157$ &$1.39$ & $1.00$ & $0.0$ \\
Rb$^b$ & & $1.00$ & $0.997$ & $0.042$  & $1.22$ & $0.999$ & $0.041$ & 
$1.41$ & $1.00$ & $0.0$ \\
K & $430.4$ & $1.00$ & $0.996$ & $0.013$ & $1.19$ & $1.000$ & 
$0.089$ & $1.42$ & $1.00$ & $0.0$ \\
Li & $127.7$ & $0.99$  & $1.003$ & $0.144$ & $1.10$ & $1.003$ & 
$0.161$ & $1.41$ & $1.00$ & $0.0$ \\
Sr & $319.6$ & $1.00$ & $1.000$ & $0.000$ & $1.21$ & $1.009$ & 
$0.383$ & $1.42$ & $1.014$ & $0.109$ \\
\hline \hline
$^a$ This work.\\
$^b$ From Ref.\cite{7}.\\
\end{tabular} \\[2mm]
\end{center}\vspace*{-1ex}
\end{figure}

\begin{multicols}{2}

\end{multicols}
\begin{figure}
\refstepcounter{table}
\small{TABLE II. Elastic constants of stable and metastable phases: The 
tetragonal elastic stiffness constants $\bar{c}_{11}$, $\bar{c}_{13}$, 
$\bar{c}_{33}$, in Mbar for phase $1$ at stationary point $1$
of Table I and for phase 2 at stationary point 3 of Table I, 
$c_{11}$, $c_{13}$, $c_{33}$ are the usual elastic constants for tetragonal 
structures. For cubic phases $c_{11} = c_{33} = \bar{c}_{33}$, hence from 
Eq.(2), $c_{12}$ is found in two ways: 
$c_{12}^{(1)} = (\bar{c}_{11}/2)-\bar{c}_{33}$ and $c_{12}^{(2)} = 
\bar{c}_{13}/2$. Note that $1$ mRy/bohr$^3=0.14711$ Mbar.}

\begin{center}
\begin{tabular}{ccccccccccccc}
\hline \hline
Metal & $c/a$  & $c_{11}$ & $c_{13}$ & $c_{33}$ & $c_{12}^{(1)}$ & 
$c_{12}^{(2)}$ &  $c/a$  & $c_{11}$ & $c_{13}$ & $c_{33}$ & $c_{12}^{(1)}$ &
$c_{12}^{(2)}$ \\
\hline 
V$^a$ & $1.00$ & $10.26$  & $3.27$  & $3.60$ & $1.53$ & $1.63$  & $1.78$ & $11.47$ & $2.08$  & $4.57$ &        & \\
V$^b$ &        &          &         & $2.88$ & $1.36$ &         &        &         &         &        &        &   \\
Ti$^a$& $0.85$ & $5.05$   & $2.18$  & $1.45$ &        &         & $1.40$ & $5.52$  & $2.38$  & $1.59$ & $1.17$ & $1.19$  \\
Ti$^b$&        &          &         &        &        &         &        &         &         & $1.48$ & $1.21$ &       \\
Rb$^a$& $0.99$ & $0.160$  & $0.074$ & $0.043$& $0.037$& $0.037$ & $1.39$ & $0.163$ & $0.075$ & $0.048$& $0.034$& $0.038$ \\
Rb$^b$&        &          &         & $0.045$& $0.038$ &         &        &         &         & $0.046$& $0.040$&        \\
K$^a$ & $1.00$ & $0.220$  & $0.100$ & $0.059$& $0.051$& $0.048$ & $1.42$ & $0.214$ & $0.097$ & $0.062$& $0.046$& $0.049$ \\
K$^b$ &        &          &         & $0.061$& $0.049$&         &        &         &         & $0.057$& $0.050$& \\
Li$^a$& $0.99$& $0.633$   & $0.296$ & $0.165$& $0.151$& $0.148$ & $1.41$ & $0.636$ & $0.280$ & $0.178$& $0.140$& $0.140$ \\
Li$^b$&       &           &         & $0.170$& $0.147$&         &        &         &         & $0.169$&  $0.143$&  \\
Sr$^a$& $1.00$& $0.678$   & $0.283$ & $0.194$& $0.145$& $0.141$ & $1.42$ & $0.698$ & $0.295$ & $0.202$&  $0.147$ & $0.147$ \\
Sr$^b$&       &           &         & $0.206$& $0.136$&         &        &         &         &  $0.189$ & $0.153$ & \\
\hline \hline
$^a$ This work.\\
$^b$ From Ref.\cite{3}.\\
\end{tabular} \\[2mm]
\end{center}\vspace*{-1ex}
\end{figure}

\begin{multicols}{2}
Despite the deviations from experiment, which may be reduced in
subsequent calculations by more accurate formulation of the electronic
structure equations, these results are of immediate practical value in
interpreting measured film structures, and of conceptual value in
providing a sharp distinction between, for example, a tetragonally 
strained bcc phase and a tetragonally strained fcc phase.  A
generalization of the tetragonal results to other structures suggests
a new type ofphase diagram in which the various equilibrium phases,
stable and metastable, are points in a parameter space which has
structural parameters as coordinates.  The present results suggest
that each phase point is surrounded by a region of strained states and
the regions are embedded in and separated by a continuous matrix 
of inherently unstable states that cannot be stabilized by applied
stresses.  Such a generalization for structures with considerable
symmetry, such as the tetragonal structure,
seems calculable by present codes.

\acknowledgements
Thanks are due to M. Scheffler of the Fritz-Haber Institute
for discussion and encouragement and to
K. Schwarz and P. Mohn of the Technical University
of Vienna for information about their energy calculations.

\end{multicols}


\begin{references}
\bibitem{6}
The relations between the $\bar{c}_{ij}$ and the $c_{ij}$ in
Eq.(\ref{Eq2}) are derived by comparing second derivatives of
$E$ obtained in two ways.  One way uses Eq.(\ref{Eq2}) and treats $E$
as a function of the two variables $a$ and $c$.  The second way treats
$E$ as a function of three orthogonal variables $a_1$ and
$a_2$ the two in-plane lattice constants, and $c$, the out-of-plane 
lattice constant, and takes second derivatives with respect to 
$a_1$ or $a_2$ or the mixed second derivative with respect to $a_1$
and to $c$, all at $a_1 = a_2 = a$.  The second way generates the
usual $c_{ij}$.
\end{references}
\end{document}